\documentclass[fleqn,12pt,twoside]{article}

\usepackage[]{espcrc1}
\usepackage{bbm}
\usepackage{graphicx}
\usepackage{amssymb, amsmath}
\usepackage{floatflt}

\title{Baryogenesis in the MSSM, nMSSM and NMSSM
\footnote{
Presented by M.G.~Schmidt at the SEWM06, Brookhaven National Laboratory, May 10-13, 2006
}
}
\author{S. J. Huber\address{
Theory Division, CERN,
CH-1211 Geneva 23, Switzerland
}, 
T. Konstandin\address{
Department of Physics, Royal Institute of Technology (KTH), 
AlbaNova University Center, 
Roslagstullsbacken 11, 106 91 Stockholm, Sweden
}, 
T. Prokopec\address{
Institute for Theoretical Physics (ITF) \& Spinoza Institute,
             Utrecht University, Leuvenlaan~4, Postbus 80.195,
             3508 TD Utrecht, The Netherlands
}, 
M.G.~Schmidt\address{
Institut f\"ur Theoretische Physik, Heidelberg
University,
Philosophenweg 16, D-69120 Heidelberg, Germany
}}

\newcommand{\nc}{\newcommand}

\nc{\be}{\begin{equation}}
\nc{\ee}{\end{equation}}
\nc{\bea}{\begin{eqnarray}}
\nc{\eea}{\end{eqnarray}}
\nc{\nn}{\nonumber}

\nc{\markx}{$\clubsuit$}

\def\Slash#1{#1\kern-0.55em\raise.05ex\hbox{/}}
\def\slash#1{#1\kern-0.5em\raise.05ex\hbox{{$\scriptstyle /$}}}

\begin{document}

\maketitle

\begin{abstract}
We compare electroweak baryogenesis in the MSSM, nMSSM and NMSSM. We
comment on the different sources of CP violation, the phase
transition and constraints from EDM measurements.
\end{abstract}

\section{Introduction to Electroweak Baryogenesis}

A viable baryogenesis mechanism aims to explain the observed asymmetry in 
the baryon density,
$\eta = \frac{n_B - n_{\bar B}}{s} \approx 8.7(3) \times 10^{-11}$, 
and the celebrated Sakharov conditions state the necessary ingredients
for baryogenesis: 
(i) C and CP violation,
(ii) non-equilibrium,
(iii) B number violation.

B number violation is present in the hot Universe due to sphaleron processes
while C is violated in the electroweak sector of the Standard Model (SM).
The two important aspects of electroweak baryogenesis
(EWBG)\cite{Cohen:1990it} are transport and CP violation. EWBG
requires a strong first-order electroweak phase transition to drive
the plasma out of equilibrium.  The CP violation is induced by the
moving phase boundary.  Hence it is important to derive transport
equations that contain CP-violating quantum effects in a genuine
manner.

Compared to other baryogenesis mechanisms, EWBG has the attractive
property that the relevant energy scale will be accessible by the next
generation of collider experiments.

\section{Transport equations derived from the Kadanoff-Baym equations}

The Kadanoff-Baym equations represent the statistical analog to the 
Schwinger-Dyson equations and are of the following form:
\be
 ( \Slash{k} + \frac{i}{2} \Slash{\partial} 
- P_L \, m    \,    e^{-\frac{i}{2} \overleftarrow{\partial} \cdot \partial_k  } 
- P_R m^\dagger \,  e^{-\frac{i}{2} \overleftarrow{\partial} \cdot \partial_k  } 
) \, 
g^< (k , X) = \textrm{collision terms}, 
\ee
and all quantities are functions of the center of mass coordinate
$X_\mu$ and the momentum $k_\mu$.  In the approximation of planar wall
profiles, the spin is conserved what can be
used to partially decouple the equations.
%

In the Kadanoff-Baym equations, the exponential derivative operator is
usually expanded which corresponds to a semi-classical expansion in
$\hbar$, the so-called gradient expansion. In the following we will
keep the first two orders in $\hbar$.


The simplest example of CP violation in transport equations is given
by the one-flavour case with a $z-$dependent complex phase in the mass
term~\cite{Kainulainen:2001cn}, $m(z)= |m(z)| \times e^{i \theta(z)}$. In
this case the constraint equation (real part of the K.-B.
equation) leads to the following {\it Ansatz} (the subscript $s$ is
the spin quantum number)
\be
g^<_s = 2\pi \, f_s \, \delta (k_0 - \omega_s), \quad
\omega_0 = \sqrt{k_z^2 + k_\parallel^2 + |m|^2}, \quad
\omega_s = \omega_0
 - s \frac{\, |m|^2 \theta^\prime}{2 \omega_0 \sqrt{\omega_0^2 - k_\parallel^2}},
\ee
where $f_s$ denotes the distribution function and $\omega_s$ the 
energy of the particle in the presence of the wall.
The kinetic equation (imaginary part of the K.-B. equation)
is of the Vlasov type 
\be
\frac{k_z}{\omega_s} \partial_z f_s + F_s \partial_{k_z} f_s = \textrm{collision terms}, \quad
F_s = -\frac{{|m|^2}^\prime}{2 \omega_s} + s \frac{(|m|^2 \theta^\prime)^\prime}
	{2 \omega_0 \sqrt{\omega_0^2 - k_\parallel^2}} \label{sem_cl_force}.
\ee
Note that the second part of
the force $F_s$ violates CP and hence sources EWBG.


The multi flavour case can be treated in the linear response
approximation, where the Green function is split according to $g^< =
g^{<,eq} + \delta g$ and leads to a kinetic equation of the
form~\cite{Konstandin:2004gy} (without using the in general
nonalgebraic constraint Eqs. that reproduce the dispersion relations
in lowest order)
\be
 k_z\partial_z \delta g 
 + \frac{i}{2}\big[m^2,\delta g\big]
 + k_0 \Gamma \delta g 
= S ( g^{<,eq} ). \nn
\ee
The third term is a damping term, taking into account the collision terms.
Note that the second term will lead to an oscillation of the
off-diagonal densities in the mass eigenbasis, similar to neutrino
oscillations. 
The right-hand side of this equation contains contributions of higher
order in the gradient expansion that give rise to CP violation and EWBG.

\section{EWBG in the MSSM}

In the MSSM the dominant contribution to baryogenesis comes from the 
charginos (Higgsino - Wino - mixing) with the mass matrix
\bea
\psi_R = \binom{\tilde W^+_L}{\tilde h_{1,R}}, \quad 
\psi_L = \binom{\tilde W^+_R}{\tilde h_{2,L}}, \quad
m(z) = \begin{pmatrix}
M_2 & g \, H_2^* (z) \\
g \, H_1^* (z) & \mu_c
\end{pmatrix},
\eea
where the SUSY breaking parameters $M_2$ and $\mu_c$ contain complex phases.

\begin{figure}[t]
\begin{center}
\includegraphics[width=8 cm]{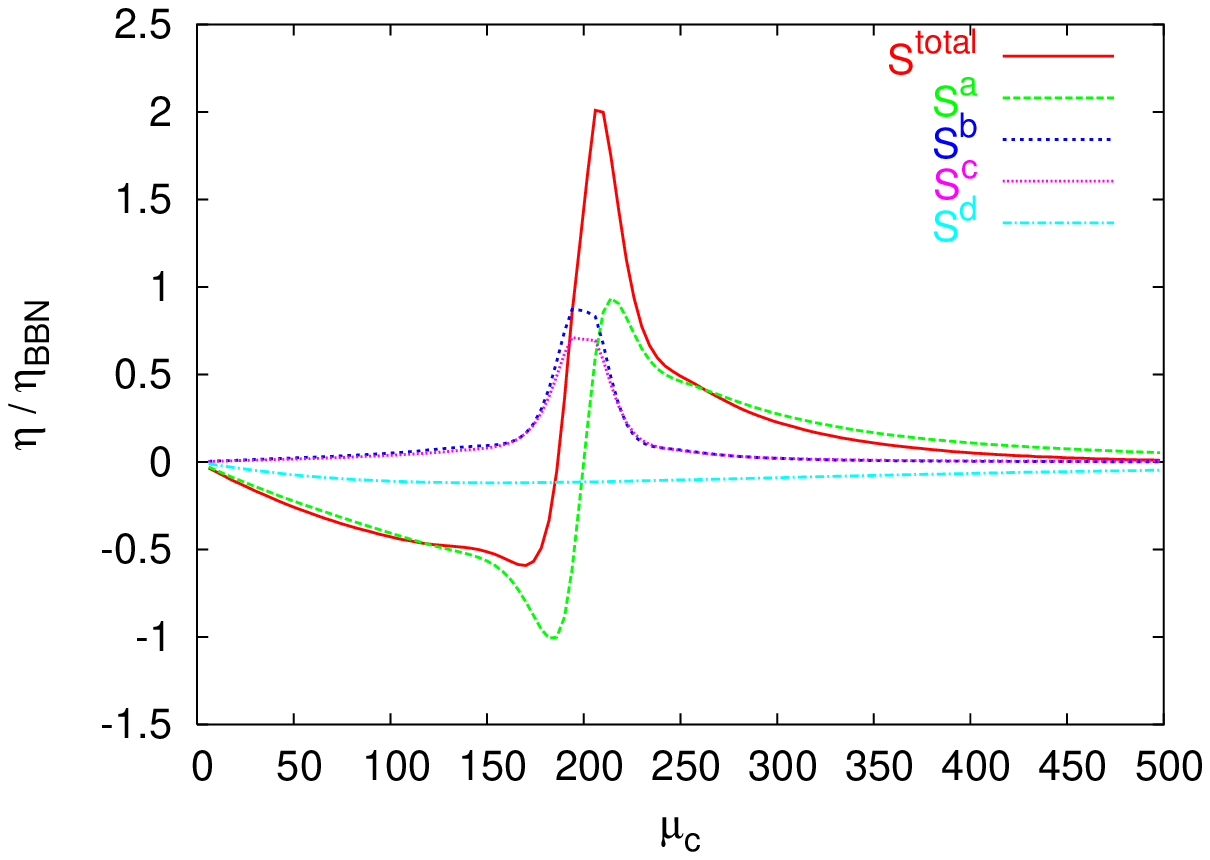}
\includegraphics[width=5.5 cm]{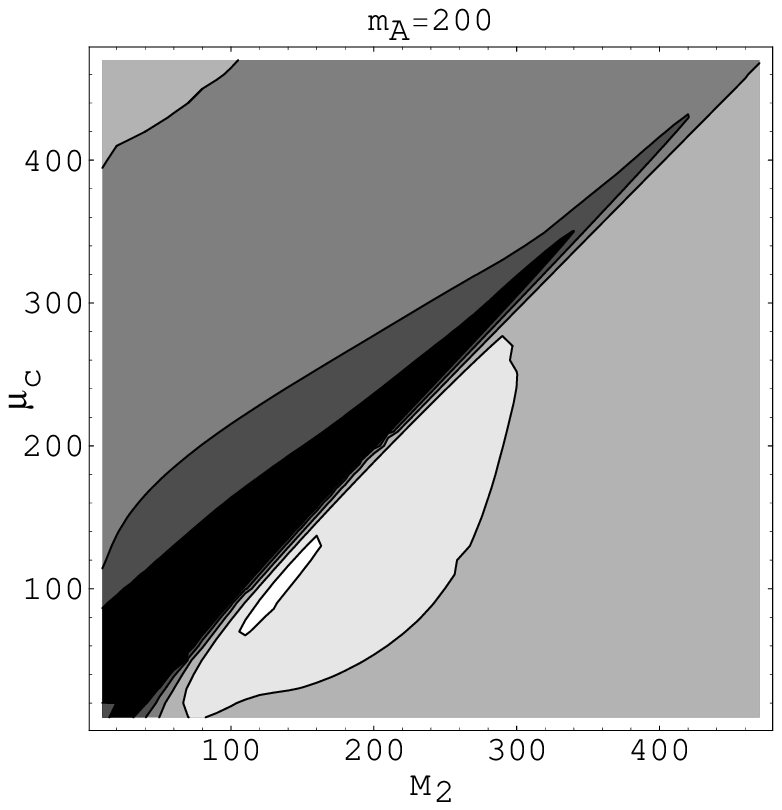}
\end{center}
\vskip -1.5 cm
\caption{ The left plot shows the produced BAU in the MSSM for $M_2=200$ GeV. 
In the right plot, the black area denotes the region in the $(\mu_c, M_2)$ plane, where
EWBG is viable.\label{MSSM_EWBG}}
\end{figure}

The CP-violating sources to first order in gradients only contribute
to the off diagonal terms in flavour space and hence are suppressed
by the oscillation effect. They read~\cite{Konstandin:2005cd} 
\bea
{\cal{S}}^a_\mu &=& 2 g^2 T_c^{-4}\Im(M_2\mu_c) (|M_2|^2 - |\mu_c|^2)
             \partial_\mu \big(H_1 H_2\big)\eta_{(0)}^3,
\nn 
\\
{\cal{S}}^b_\mu &=&  2 g^4 T_c^{-4}\Im(M_2\mu_c) (H_1^2-H_2^2)
       \partial_\mu(H_1 H_2)\eta_{(0)}^3,
\nn \\
{\cal{S}}^c_\mu  &=& -2 g^2 T_c^{-2}
      \Im(M_2\mu_c)\big( H_2\partial_\mu H_1 - H_1\partial_\mu H_2\big)
                   \big(\eta_{(0)}^0 + 4\eta_{(2)}^3\big).
\eea
In addition, there is a CP-violating source of second order in the
gradient expansion that contributes to the diagonal elements in flavour space and
corresponds to the semi-classical force that appears in the one
flavour case in Eq.~(\ref{sem_cl_force})
\be
{\cal{S}}^d_0 =  2 \, v_w\, g^2 T_c^{-4}\Im(M_2\mu_c)
                  \big(H_2\partial^2_z H_1 + H_1\partial^2_z H_2\big)
                   \zeta_{(0)}^3.
\ee
The functions $\eta$ and $\zeta$ denote certain momentum integrals of
the equilibrium distribution functions.  Fig.~\ref{MSSM_EWBG} shows
the produced baryon asymmetry for a maximal CP-violating phase in the
chargino mass matrix using the system of diffusion equations suggested
in Ref.~\cite{Huet:1995sh}.
In the right plot, the black area denotes the region of the parameter space
where EWBG is viable. 
We conclude that EWBG in the MSSM is only possible if:
(i) The charginos are nearly mass degenerate such that mixing effects 
are not suppressed.
(ii) The CP phases in the chargino sector are $O(1)$.

\section{EWBG in singlet extensions of the MSSM}

The general NMSSM of Ref.~\cite{Huber:2000mg} consists of the MSSM
extended by a gauge singlet and the superpotential
\be
W_{NMSSM} = \lambda S H_1 H_2 + \frac{k}{3} S^3 + \mu H_1 H_2 + r S + W_{MSSM}.
\ee
\begin{figure}[t]
\centering
\includegraphics[width=0.6\textwidth]{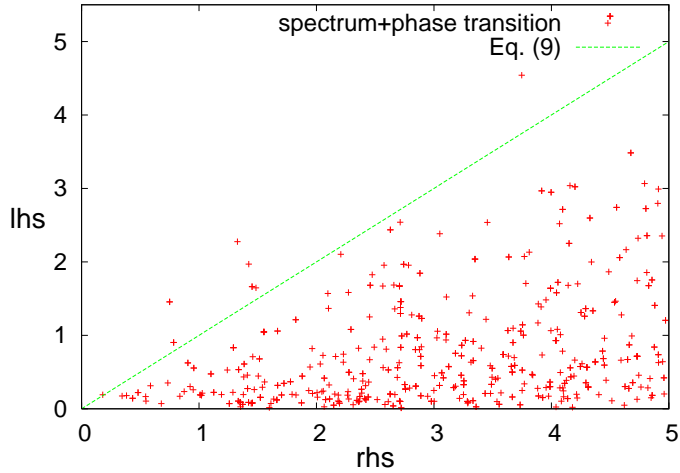}
\caption{The electroweak phase transition in the nMSSM. \label{PT_NMSSM}}
\end{figure}
Due to the explicit $\mu$ term and the singlet self-couplings, this
model provides a rich Higgs phenomenology; however, additional
assumptions have to be made to prevent higher-dimensional operators
from destabilizing the hierarchy. It does not suffer from a domain 
wall problem since there are no discrete symmetries. 

In the nMSSM, a $\mathbb{Z}_5$ or $\mathbb{Z}_7$ symmetry is imposed
to solve the domain wall problem without destabilizing the electroweak
hierarchy. The $\mu$ term is forbidden and only induced after
electroweak symmetry breaking. Thus the $\mu$ problem is solved.  The
discrete symmetries also eliminate the singlet self coupling.  A
rather large value of lambda is needed in the nMSSM to fulfill current
mass bounds on the Higgsinos and charginos, which might lead to a
Landau pole below the GUT scale.

\subsection{Electroweak phase transition}

In contrast to the MSSM, no light stop is needed in the NMSSM or
nMSSM, since the additional singlet terms in the Higgs potential
strengthen the phase transition~\cite{Huber:2000mg}. In the nMSSM
case these terms read:
\be
{\cal L} = {\cal L}_{MSSM} + m_s^2 |S|^2
        + \lambda^2 |S|^2 (H_1^\dagger H_1 + H_2^\dagger H_2)
 +\, t_s (S + \, h.c.) + (a_\lambda S H_1 \cdot H_2 + \, h.c.).
\ee
In a simplified scheme without CP violation, a first-order phase
transition due to tree-level dynamics occurs if~\cite{Menon:2004wv}
\be
m_s^2 < \frac{1}{\tilde \lambda} \left| \frac{\lambda^2 t_s}{m_s} 
	- m_s \tilde a \right|, \quad 
\tilde a = \frac{a_\lambda}{2} \, \sin 2 \beta, \quad 
\tilde \lambda^2 = \frac{\lambda^2}{4} \sin^2 2\beta + 
\frac{\bar g^2}{8} \cos^2 2\beta.
\label{PT_ineq}
\ee
Fig. \ref{PT_NMSSM} displays Eq.~(\ref{PT_ineq}) for random nMSSM
models with a strong PT and shows that this criterion is also
decisive if CP violation and the one-loop effective potential are
taken into account.

\subsection{EDM constraints and baryon asymmetry}

Since the trilinear term in the superpotential contributes to the
Higgs mass, $\tan (\beta)$ is generically of $O(1)$; Hence two-loop
contributions from the charginos to the electron EDM are naturally
small. The one-loop contributions to the electron EDM can, as in the
MSSM, be reduced by increasing the sfermion masses.

\begin{figure}[t]
\centering
\includegraphics[width=0.6\textwidth]{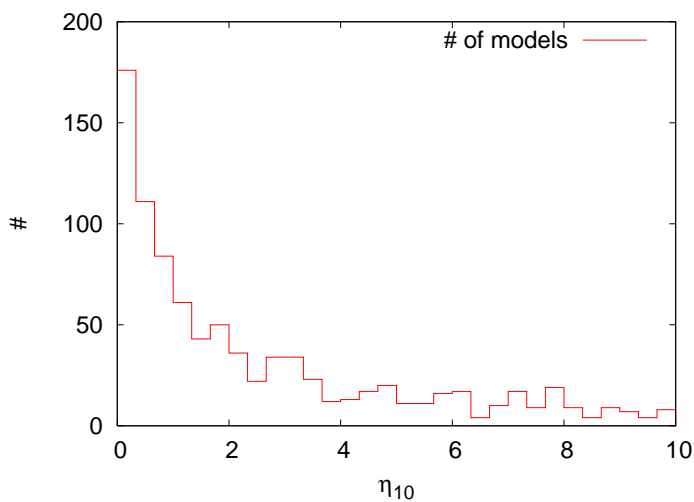}
\caption{Produced baryon asymmetry in random nMSSM models. \label{EWBG_nMSSM}}
\end{figure}
The effective $\mu$ parameter is dynamical in the nMSSM and NMSSM, and
hence its complex phase can change during the phase transition. This
leads to new CP-violating sources in the chargino sector that are of
second order in the gradient expansion and do not rely on mixing. Thus,
these contributions are not suppressed by the flavour oscillations and
mass degenerate charginos are not required for viable EWBG.

Additionally, the bubble wall tends to be thinner than in the MSSM and
hence it is rather generic to generate the observed baryon
asymmetry~\cite{Huber:2006wf}. Fig.~\ref{EWBG_nMSSM} shows the binned
BAU for a random set of nMSSM models with a strong first order PT.

Another interesting feature of the NMSSM is transitional CP violation.
If universality is violated in the singlet sector of the NMSSM, the
phase transition can connect a high temperature phase with broken CP
and a low temperature phase of conserved CP~\cite{Huber:2000mg}. Such
CP violation cannot be detected by zero temperature experiments! The
singlet self coupling is important to stabilize the singlet in this
scenario.

\section{Conclusions}
Singlet extensions of the MSSM provide a framework in which electroweak baryogenesis 
seems to be possible without fine tuning.


\begin{thebibliography}{99}
\bibliographystyle{unsrt}


\bibitem{Cohen:1990it}
  A.~G.~Cohen, D.~B.~Kaplan and A.~E.~Nelson,
  Nucl.\ Phys.\ B {\bf 349} (1991) 727.

\bibitem{Kainulainen:2001cn}
  K.~Kainulainen, T.~Prokopec, M.~G.~Schmidt and S.~Weinstock,
  JHEP {\bf 0106} (2001) 031
  [hep-ph/0105295].

\bibitem{Konstandin:2004gy}
  T.~Konstandin, T.~Prokopec and M.~G.~Schmidt,
  Nucl.\ Phys.\ B {\bf 716} (2005) 373
  [hep-ph/0410135].

\bibitem{Konstandin:2005cd}
  T.~Konstandin, T.~Prokopec, M.~G.~Schmidt and M.~Seco,
  Nucl.\ Phys.\ B {\bf 738} (2006) 1
  [hep-ph/0505103].

\bibitem{Huet:1995sh}
  P.~Huet and A.~E.~Nelson,
  Phys.\ Rev.\ D {\bf 53} (1996) 4578
  [hep-ph/9506477].

\bibitem{Panagiotakopoulos:2000wp}
  C.~Panagiotakopoulos and A.~Pilaftsis,
  Phys.\ Rev.\ D {\bf 63} (2001) 055003
  [hep-ph/0008268].

\bibitem{Huber:2000mg}
  S.~J.~Huber and M.~G.~Schmidt,
  Nucl.\ Phys.\ B {\bf 606}, 183 (2001)
  [hep-ph/0003122];
%
  S.~J.~Huber and M.~G.~Schmidt,
  Proceedings SEWM '00, Marseille, 272-278
  [hep-ph/0011059].

\bibitem{Menon:2004wv}
  A.~Menon, D.~E.~Morrissey and C.~E.~M.~Wagner,
  Phys.\ Rev.\ D {\bf 70} (2004) 035005
  [hep-ph/0404184].

\bibitem{Huber:2006wf}
  S.~J.~Huber, T.~Konstandin, T.~Prokopec and M.~G.~Schmidt,
  hep-ph/0606298.


\end{thebibliography}
\end{document}